\documentclass{elsart}

\newcounter{bla}

\newcommand{\bea}{\begin{eqnarray}}
\newcommand{\eea}{\end{eqnarray}\noindent}
\newcommand{\nn}{\nonumber}
\newcommand{\calst}{\mbox{$\cal S$}}
\newcommand{\icalst}[1]{\mbox{${\cal S}^{-1}_{#1}$}}

\def\eps{\epsilon}
\usepackage{amssymb}
\usepackage{epsfig}

\begin{document}

\bibliographystyle{unsrt}

\begin{frontmatter}

\hfill{Edinburgh 2008/40}\\
\hfill{LAPTH-1277/08}\\
\hfill{IPPP/08/73, DCPT/08/146}\\ 
\hfill{Nikhef-2008-26}

\title{Golem95: a numerical program to calculate one-loop tensor integrals
with up to six external legs}

\author[a]{T.Binoth},
\author[b]{J.-Ph.Guillet},
\author[c]{G.Heinrich},
\author[b]{E.Pilon},
\author[d]{T.Reiter}


\address[a]{School of Physics, The University of Edinburgh, Edinburgh EH9 3JZ, UK}
\address[b]{LAPTH, Universit\'e de Savoie and CNRS, Annecy-le-Vieux, France}
\address[c]{Institute for Particle Physics Phenomenology,
        University of Durham, \\Durham, DH1 3LE, UK}
\address[d]{NIKHEF, Kruislaan 409, 1098 SJ Amsterdam, The Netherlands}
	

\begin{abstract}
We present a program for the numerical evaluation of form factors 
entering the calculation of one-loop amplitudes 
with up to six external legs. The program is written in Fortran95 and 
performs the reduction 
to a certain set of basis integrals numerically, 
using a formalism where inverse Gram determinants can be avoided.
It can be used to calculate one-loop amplitudes with massless internal 
particles in a fast and numerically stable way. 

\begin{flushleft}
\end{flushleft}


\end{abstract}

\end{frontmatter}

{\bf PROGRAM SUMMARY}

\begin{small}
\noindent
{\em Manuscript Title:} Golem95: a numerical program to calculate one-loop diagrams 
with up to six external legs  \\
{\em Authors:} T.~Binoth, J.-Ph.~Guillet,  G.~Heinrich, E.~Pilon, T.~Reiter\\
{\em Program Title: } golem95\_v1.0                                          \\
{\em Journal Reference: }                                      \\
{\em Catalogue identifier:}                                   \\
{\em Licensing provisions:} none    \\
{\em Programming language:} Fortran95   \\
{\em Computer: Any computer with a  Fortran95 compiler }   \\
{\em Operating system:} Linux, Unix                                        \\
{\em RAM:} RAM used per form factor  is insignificant, even for a  rank six six-point form factor\\
{\em Number of processors used:} one                              
{\em Classification:} 4.4 Feynman diagrams, 11.1 High Energy Physics Computing   \\
{\em External routines/libraries:} perl                                      \\

{\em Nature of problem:} Evaluation of 
one-loop multi-leg tensor integrals occurring in the calculation of next-to-leading order corrections
to scattering amplitudes in elementary particle physics.\\
   \\
{\em Solution method:} Tensor integrals are represented in terms of form factors 
and a set of basic building blocks (``basis integrals"). 
The reduction to the basis integrals is performed numerically, 
thus avoiding the generation of large algebraic expressions.
\\
   \\
{\em Restrictions:} The current version contains basis integrals for 
massless internal particles only.
Basis integrals for massive internal particles will be 
included in a future version.
\\
{\em Running time: } Depends on the nature of the problem. 
A rank 6 six-point form factor 
at a randomly chosen kinematic point takes 0.13 seconds on an 
Intel Core 2 Q9450 2.66\,GHz processor.

\end{small}

\newpage

\hspace{1pc}
{\bf LONG WRITE-UP}

\section{Introduction}

Collider experiments at the TeV scale, in particular
the LHC experiments,  are 
 expected to shed light on the mechanism of electroweak symmetry breaking 
and to open up new horizons concerning our understanding of elementary particle 
interactions. 
In order to achieve these goals, expected signal as well as background rates 
should be well under control, which implies 
that a multitude of scattering processes should be 
known at next-to-leading order (NLO) accuracy. 

Over the last years, enormous efforts have been made to calculate NLO corrections, 
in QCD as well as in the electroweak sector. For a review 
see e.g. \cite{Bern:2008ef}.
These calculations in general involve two parts, the treatment of extra real emission 
and the calculation of virtual corrections, i.e. one-loop amplitudes.
While the calculation of one-loop amplitudes with up to four external particles 
has reached a quite mature state meanwhile, and automated tools 
have been developed already some time ago~\cite{vanOldenborgh:1989wn,Mertig:1990an,Hahn:1998yk,Yuasa:1999rg,Hahn:2000kx}, 
the calculation of processes with five or more external legs 
required and boosted new developments in various directions, 
for recent developments see 
e.g. \cite{Bern:2008ef,Binoth:2005ff,Denner:2005fg,Denner:2005nn,Ellis:2005zh,Ossola:2006us,Binoth:2006hk,Anastasiou:2006gt,Bern:2007dw,Ellis:2007br,Kilgore:2007qr,Britto:2008vq,Britto:2008sw,Giele:2008ve,Mastrolia:2008jb,Catani:2008xa,Ellis:2008ir,Glover:2008ffa}.

Initially, NLO calculations have mostly been done on a process-by-process basis, 
but fortunately we are moving  towards automation  also 
for multi-particle processes, as can be seen from the  
tools which have been constructed recently. 
For the automated calculation of one-loop amplitudes with more 
than four external legs, there are 
the publicly available programs 
{\tt FormCalc/Loop\-Tools}\,\cite{Hahn:1998yk,Hahn:2000kx}  which recently have been extended to 
  5-point processes \cite{Hahn:2006ig,Hahn:2006qw} and 
the program {\tt CutTools} \cite{Ossola:2007ax} which is 
based on a numerical unitarity formalism~\cite{Ossola:2006us,Ellis:2007br,Giele:2008ve,Mastrolia:2008jb,Ellis:2008ir}.
Further, there are the programs 
{\tt BlackHat}~\cite{Berger:2008sj} and {\tt Rocket}~\cite{Giele:2008bc}, 
relying also on cutting techniques. 
Concerning the generation of subtraction terms for real radiation, 
automated tools have become publicly available recently as well~\cite{Gleisberg:2007md,Seymour:2008mu,Hasegawa:2008ae,Frederix:2008hu}.
Integral libraries for massive~\cite{vanOldenborgh:1989wn,Hahn:2006qw} as well as 
infrared divergent~\cite{Ellis:2007qk} scalar
integrals also exist.

As already mentioned, a public program for the reduction of tensor integrals so far is available only for 
infrared-finite integrals and for up to five external legs~\cite{Hahn:1998yk,Hahn:2006qw}. 
In this paper we present a program for the numerical reduction of tensor integrals 
with up to six external legs.
In the present version, we focus mainly on massless QCD applications, 
i.e. processes with massless internal particles. 
The  master integrals are implemented in the code
to be valid in all kinematic regions. Infrared divergences are 
regulated dimensionally, i.e  the loop momenta live in $n=4-2\epsilon$ dimensions. 
The output for a specific kinematic point 
is a set of six numbers representing the real and imaginary parts of the 
coefficients of the Laurent series in  $\epsilon$, 
i.e. the coefficients of the $1/\epsilon^2, 1/\epsilon$ poles and the finite part.

The reduction formalism is valid for massless as well as massive external and internal 
particles. However, the basis integrals for processes involving {\it internal} massive
particles will be implemented  in a forthcoming version.
We would like to emphasize that the program can be used not only for tensor reduction, 
but also to calculate basis integrals, with or without Feynman parameters in the numerator, 
and therefore is also of interest for calculations where the integral coefficients have
been determined by unitarity cut techniques: {\tt golem95} can be used as a library for 
master integrals.

The paper is organised as follows. In section 2, we review shortly the theoretical background. 
Section 3 contains a brief summary of the software structure, while section 4 contains a detailed
description of the individual components of the program.
The installation instructions are given in section 5, and section 6 contains the descriptions 
of three different test runs: 
the calculation of a form factor for a rank five five-point function, 
the calculation of all form factors for rank one six-point functions in one go, 
and finally the calculation of a full amplitude: the helicity amplitudes for 
light-by-light scattering. 
We give an outlook on future versions in section 7 and explain technical details in appendices 
\ref{landau} and \ref{onedimint}.
The code comes with a number of demonstration programs, demonstrating for example 
the behaviour near a scattering singularity, or the relation to LoopTools notation.
All these demonstration programs are listed in Appendix \ref{demos}.

\section{Theoretical background}

The program is an implementation of the formalism developed in Ref.~\cite{Binoth:2005ff}.
Here we will summarize only its main features relating to the {\tt golem95} program, 
for further details we refer to \cite{Binoth:2005ff}.

\subsection{Form Factors}

\begin{figure}[ht]
\unitlength=1mm
\begin{picture}(150,60)
\put(55,5){\includegraphics[width=5cm, height=5cm]{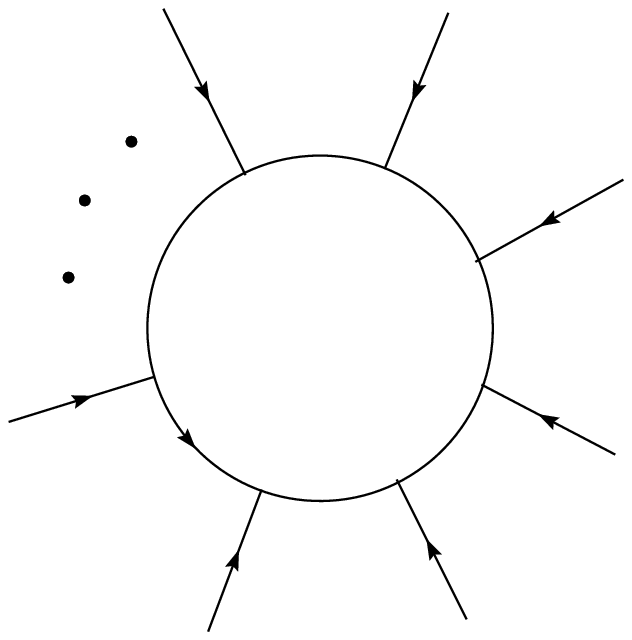}}
\put(50,15){$p_{N-2}$}
\put(60,5){$p_{N-1}$}
\put(85,5){$p_{N}$}
\put(100,15){$p_{1}$}
\put(102,34){$p_{2}$}
\put(92,50){$p_{3}$}
\put(70, 54){$p_{4}$}
\put(87,21){\footnotesize $N$}
\put(90,28){\footnotesize $1$}
\put(87,35){\footnotesize $2$}
\put(79,38){\footnotesize $3$}
\end{picture}
\caption{General $N$-point one-loop graph with momentum and propagator labelling.}
\label{fig1}
\end{figure}
A general one-loop tensor integral of rank $r$ can be written as 
\begin{eqnarray}
I^{n,\,\mu_1\ldots\mu_r}_N(a_1,\ldots,a_r) = 
\int \frac{d^n k}{i \, \pi^{n/2}}
\; \frac{q_{a_1}^{\mu_1}\,\dots  q_{a_r}^{\mu_r}}{
(q_1^2-m_1^2+i\delta)\dots (q_N^2-m_N^2+i\delta)}
\label{eq0}
\end{eqnarray} 
where $q_a=k+r_a$, and $r_a$ is a combination of external momenta. 
For the diagram in Fig.~\ref{fig1}, $r_i=\sum_{j=1}^i p_j$.
Our method is defined in $n=4-2\epsilon$ dimensions and thus is
applicable to general scattering processes with arbitrary 
propagator masses.
Taking integrals of the form (\ref{eq0}), i.e. with $q_{a}^{\mu}$ 
instead of just $k^\mu$ in the numerator, 
as building blocks has two advantages: first, combinations of
loop and external momenta appear naturally  in Feynman rules,   
second, it allows for a formulation of the tensor reduction which 
manifestly maintains 
the invariance of the integral under a shift $k \to k+r_0$  in the loop momentum. 
Such a shift can be
absorbed into a redefinition of the $r_j, \,r_j \to r_j - r_0$. 
By setting $a_1,\ldots, a_r=N$, and using momentum conservation to set $r_N=0$, 
we can always retrieve the commonly used form 
\begin{equation}
I^{n,\,\mu_1\ldots\mu_r}_N(N,\dots,N) = 
\int \frac{d^n k}{i \, \pi^{n/2}}\,
\frac{k^{\mu_1}\ldots k^{\mu_r}}{(q_1^2-m_1^2+i\delta)\dots (q_N^2-m_N^2+i\delta)}\;.
\label{conventional}
\end{equation}

The Lorentz structure of the integral (\ref{eq0}) is carried by
tensor products of the  metric $g^{\mu\nu}$ and the difference vectors 
\begin{equation}
\Delta_{ij}^\mu=
r_i^\mu - r_j^\mu\;,
\end{equation}
 which are shift invariant. 
Therefore, tensor integrals are expressible by linear combinations 
of such Lorentz tensors and 
{\it form factors} $A^{N,r}_{l_1\cdots l_r}$,
$B^{N,r}_{l_1\cdots }$, $C^{N,r}_{\cdots }$, 
defined by 
\bea 
\lefteqn{I^{n,\,\mu_1\ldots\mu_r}_N(a_1,\ldots,a_r;\,S) =}  
\nn \\
& &  
\; 
\sum_{j_1\cdots j_{r}\in S} \;\;\;
\left[ 
 \Delta_{j_1\cdot}^{\cdot} \cdots \Delta_{j_r\cdot}^{\cdot} 
\right]^{\{\mu_1\cdots\mu_r\}}_{\{a_1\cdots a_r\}} \, A_{j_1 \ldots ,j_{r}}^{N,r}(S) 
\nn\\ 
&+& 
\sum_{j_1\cdots j_{r-2}\in S} \, 
\left[
 g^{\cdot\cdot} \Delta_{j_1\cdot}^{\cdot} \cdots \Delta_{j_{r-2}\cdot}^{\cdot}
 \right]^{\{\mu_1\cdots\mu_r\}}_{\{a_1\cdots a_{r}\}}\,  B_{j_1 \ldots,j_{r-2}}^{N,r}(S) 
\nn\\ 
&+& 
\sum_{j_1\cdots j_{r-4}\in S} \, 
\left[
 g^{\cdot\cdot}g^{\cdot\cdot}  \Delta_{j_1\cdot}^{\cdot} \cdots
                               \Delta_{j_{r-4}\cdot}^{\cdot} 
\right]^{\{\mu_1\cdots\mu_r\}}_{\{a_1\cdots a_{r}\}}\, C_{j_1 \ldots ,j_{r-4}}^{N,r}(S) 
\label{fofageneral} 
\eea
where $[\cdots]^{\{\mu_1\cdots\mu_r\}}_{\{a_1\cdots a_r\}}$ 
denotes the distribution of the $r$ Lorentz indices $\mu_i$, and momentum 
labels $a_i$ to the vectors $\Delta_{j\,a_i}^{\mu_i}$ in all distinguishable ways. 
$S$ denotes an ordered 
set of propagator labels, related to 
the kinematic matrix ${\cal S}$, defined by 
\bea
\calst_{ij} &=&  (r_i-r_j)^2-m_i^2-m_j^2\;\quad ; \;\quad i,j\in\{1,\ldots,N\}\;.
\label{eqDEFS}
\eea
There is a one-to-one correspondence between $\calst_{ij}$ 
and the set $S=\{1,\ldots,N\}$.
We recall that  standard form factor representations can be simply obtained
by replacing  $a_j=N$ for all $j$, together with $r_N=0$.
This also shows that the form factors do {\em not} depend on the introduction
of the difference vectors $\Delta_{ij}^\mu$. 
The form factors are shift invariant by themselves.
Therefore the program {\tt golem95} can be used without ever introducing 
difference vectors, if the user prefers not to do so. 

Due to the fact that for $N\geq 5$, 
four linearly independent external vectors form a basis
of Minkowski space, the tensor reduction for $N\geq 6$ 
can be done in such a way that only  form factors for $N\le 5$ 
are needed.
Therefore, the  Lorentz structure of $(N>5)$-point rank $r$ tensor integrals
does not require the introduction of additional factors of $g^{\mu\nu}$ 
as compared to the $N=5$ case, only additional external vectors  
appear.  We note that for $N=5$, 
one could already  express 
the metric by external momenta, but this would introduce 
inverse Gram determinants.
In \cite{Bern:1993kr}, it is  shown that all tensor 
five-point functions can 
be reduced to some basis integrals  without generating higher dimensional five-point 
functions.  In \cite{Binoth:2005ff}, a formal proof of this fact can be found, 
as well as a reduction method which avoids both inverse Gram determinants 
and spurious higher dimensional five-point functions.  
A method where inverse Gram determinants 
in the reduction from five-point to four-point integrals are absent is also presented in
Ref. \cite{Denner:2005nn}.

The form factors are linear combinations of reduction coefficients and 
basis\footnote{We call them ``basis integrals "  because 
they are the endpoints of our reduction, although
they do not form a basis in the mathematical sense.} 
integrals, where our basis integrals are not necessarily scalar 
integrals, as explained in section \ref{basisints}. 
The reduction coefficients are derived from the kinematic matrices 
${\cal S}$, where we define 
\bea
b_i&=&\sum_{k\in S}{\cal S}_{ki}^{-1}\;,\; B=\sum_{i \in S} b_i \;.
\label{bi}
\eea
The quantity $B$ is related to the Gram determinant by 
\begin{equation}
B\;\det{\cal S}= (-1)^{N+1}\det G\;.\label{sumB}
\end{equation} 
The form factors are all given explicitly in \cite{Binoth:2005ff}.
As an example, a rank two pentagon integral is represented as
\bea
I_5^{n,\mu_1 \mu_2}(a_1,a_2;S) 
&=&  
\sum_{l_1,l_2 \in S}  \;
\Delta^{\mu_1}_{l_1 \, a_1} \;  \Delta^{\mu_2}_{l_2 \, a_2} \, 
A^{5,2}_{l_1 \, l_2}(S) + g^{\mu_1 \, \mu_2} \, B^{5,2}(S)  
\label{eqNpr2}\\ 
&&\nn\\ 
B^{5,2}(S)&=&- \frac{1}{2} \, \sum_{j \in S} \, b_j \, 
I_4^{n+2}(S\setminus\{j\})\nn\\
A^{5,2}_{l_1 \, l_2}(S)&=&\sum_{j \in S} \, 
\left( \, \icalst{j \, l_1} \, b_{l_2} + \icalst{j \, l_2} \, 
b_{l_1} - 2 \, \icalst{l_1 \, l_2} \, b_{j} + b_{j} \, 
{\cal S}^{\{j\}-1}_{l_1 \, l_2}  \right) \, I^{n+2}_4(S\setminus\{j\}) \nonumber \\
& & \mbox{} + \frac{1}{2} \, \sum_{j \in S} \, 
\sum_{k \in S\setminus\{j\}} \, \left[ \icalst{j \, l_2} \, 
{\cal S}^{\{j\}-1}_{k \, l_1}  + \icalst{j \, l_1} \, {\cal S}^{\{j\}-1}_{k \, l_2}  
\right]
I_3^{n}(S\setminus\{j,k\}) \nn
\eea
and it is the form factors like $A^{5,2}_{l_1\,l_2}(S), B^{5,2}(S)$ which are implemented in 
{\tt golem95}.
 
The program {\tt golem95} can be used for amplitude calculations in several ways: 
One approach, which aims to avoid tensor integrals of high rank, 
is to cancel the reducible numerators of an expression before 
interfacing  to {\tt golem95} to calculate the irreducible tensor integrals. 
However, as all form factors for maximal rank (in a renormalisable gauge)
are implemented, the expression for an amplitude can also be 
interfaced to {\tt golem95} without performing any cancellations
between numerators and propagators. This has the advantage that 
the algebraic manipulations to do are minimal, and that even the 
Dirac traces, which often appear as coefficients 
of the form factors, can be done numerically. 

\subsection{Feynman parameter representations}\label{Feynpar}

The {\tt golem95} program uses the fact that tensor integrals 
are related to  Feynman parameter integrals with Feynman parameters in the numerator.
The basic object is the set $S$, containing the labels of the propagators 
which define the integral. 
A scalar integral, after Feynman parametrisation, can be written as 
\bea
I^n_N(S) &=& (-1)^N\Gamma(N-\frac{n}{2})\int \prod_{i=1}^N dz_i\,
\delta(1-\sum_{l=1}^N z_l)\,\left(R^2\right)^{\frac{n}{2}-N}\nn\\
&& R^2 =  
-\frac{1}{2} \sum\limits_{i,j=1}^N z_i\,\calst_{ij}  z_j\,\,-i\delta
\;.
\label{isca2}
\eea
In general, a one-loop $N$-point amplitude will contain  
$N$-point integrals as well as $(N-1),(N-2),\ldots, (N-M)$-point integrals
with tree graphs attached to some of the external legs of the loop integral. 
The latter are characterised by the omission (``pinch") of some propagators 
(say $j_1,\ldots,j_m$) of the ``maximal" one loop $N$-point graph, 
and therefore correspond to a subset of $S$ where certain 
propagator labels are missing, 
$S\setminus\{j_1,\ldots,j_m\}$.  The program {\tt golem95} is based on this concept of 
sets characterising the integrals.

The general relation between tensor integrals and parameter integrals 
with Feynman parameters in the numerator is
 well known~\cite{Davydychev:1991va,Tarasov:1996br,Bern:1992em,Binoth:1999sp}
\bea
&&I^{n,\,\mu_1\ldots\mu_r}_N(a_1,\ldots,a_r\,;S) 
 =  
(-1)^r \sum_{m=0}^{[r/2]} \left( -\frac{1}{2} \right)^m\nn\\ 
&&
\sum_{j_1\cdots j_{r-2m}=1}^N \left[ 
 (g^{..})^{\otimes m}\,\Delta_{j_1\cdot}^{\cdot} \cdots \Delta_{j_r\cdot}^{\cdot}
\right]^{\{\mu_1\cdots\mu_r\}}_{\{a_1\cdots a_r\}}
\;
 I_N^{n+2m}(j_1 \ldots ,j_{r-2m}\,;S)\;,
\label{eq32}
\eea 
where $I_N^{n+2m}(j_1 \ldots ,j_{r-2m}\,;S)$ 
 is an integral with Feynman parameters in the numerator.
 $[r/2]$ stands for the nearest integer less or equal to $r/2$ and the symbol 
 $\otimes m$ indicates
 that $m$ powers of the metric tensor are present.
Feynman parameter integrals corresponding to 
diagrams where propagators $j_1,\dots,j_m$ are pinched  
with respect to the ``maximal" topology 
can be defined as
\bea
&&I^n_N(j_1,\dots,j_r;S\setminus \{l_1,\dots,l_m\}) =(-1)^N\Gamma(N-\frac{n}{2})
\nn\\
&& 
\int \prod_{i=1}^N dz_i\,
\delta(1-\sum_{k=1}^N z_k)\,
\delta(z_{l_1})\dots \delta(z_{l_m})z_{j_1}\dots z_{j_r}\left(R^2\right)^{n/2-N}\;.
\label{isca_pinch}
\eea

\subsection{Basis integrals}\label{basisints}

The basis integrals, i.e. the endpoints of our reduction, 
are  4-point functions in 6 dimensions
$I_4^6$, which are IR and UV finite, UV divergent 4-point functions in
$n+4$ dimensions, and various 2-point and 3-point functions, some of
the latter  with Feynman parameters in the numerator. This provides us with a very
convenient  separation of IR/UV divergences, as the IR poles are 
exclusively contained in
the triangle functions. Explicitly, our reduction 
basis is given by integrals of the type 
\begin{eqnarray}\label{basis_integral}
I^{n}_3(j_1, \ldots ,j_r) &=& 
-\Gamma \left(3-\frac{n}{2} \right) \, \int_{0}^{1} 
\prod_{i=1}^{3} \, d z_i \, \delta(1-\sum_{l=1}^{3} z_l) 
\, \frac{z_{j_1} \ldots z_{j_r}}{ 
(-\frac{1}{2}\, z \cdot \calst
\cdot z-i\delta)^{3-n/2}}\;,\nn\\
I^{n+2}_3(j_1) &=& 
-\Gamma \left(2-\frac{n}{2} \right) \, \int_{0}^{1} 
\prod_{i=1}^{3} \, d z_i \, \delta(1-\sum_{l=1}^{3} z_l) 
\, \frac{z_{j_1}}{ 
(-\frac{1}{2}\, z \cdot \calst
\cdot z-i\delta)^{2-n/2}}\;,\nn\\
I^{n+2}_4(j_1, \ldots ,j_r) &=& 
\Gamma \left(3-\frac{n}{2} \right) \, \int_{0}^{1} 
\prod_{i=1}^{4} \, d z_i \, \delta(1-\sum_{l=1}^{4} z_l) 
\, \frac{z_{j_1} \ldots z_{j_r}}{ 
(-\frac{1}{2}\, z \cdot \calst
\cdot z-i\delta)^{3-n/2}}\;,\nn\\
I^{n+4}_4(j_1) &=& 
\Gamma \left(2-\frac{n}{2} \right) \, \int_{0}^{1} 
\prod_{i=1}^{4} \, d z_i \, \delta(1-\sum_{l=1}^{4} z_l) 
\, \frac{z_{j_1}}{ 
(-\frac{1}{2}\, z \cdot \calst
\cdot z-i\delta)^{2-n/2}}\;,\nn\\
\end{eqnarray}
where $r^{\rm{max}}=3$, as well as $I^{n}_3,I^{n+2}_3,I^{n+2}_4,I^{n+4}_4$ 
with no Feynman parameters in the numerator, 
and two-point functions.

Note that $I^{n+2}_3$ and $I^{n+4}_4$ are UV divergent, while  $I^{n}_3$
can be IR divergent. In the code, the integrals are represented as 
arrays containing the coefficients of their Laurent expansion in $\epsilon=(4-n)/2$.

Further reduction of these integrals to scalar basis integrals (i.e. integrals with 
no Feynman parameters in the numerator)  introduces factors of $1/B$, i.e. 
inverse Gram determinants. A particular feature of {\tt golem95} is the fact that 
the above integrals are {\it not} reduced to scalar basis integrals
in cases where $B$ becomes small, thus avoiding problems with small inverse determinants.
In these cases, the above integrals are evaluated numerically. 
As $B= (-1)^{N+1}\det(G)/\det({\cal S})$ is a dimensionful quantity, 
the switch to the numerical evaluation of the basis integrals is 
implemented such that the value of the dimensionless parameter $\hat{B}$ is tested, where 
\begin{equation}
\hat{B}=B\times (\rm{largest \; entry \; of \;} {\cal S})\;.
\label{bhat}
\end{equation}
If $\hat{B}>\hat{B}^{\rm{cut}}$, the reduction is performed, else the program 
switches to the direct numerical evaluation of the integral.
The default value is $\hat{B}^{\rm{cut}}=0.005$.
A major improvement with respect to the numerical evaluation method 
used in \cite{Binoth:2005ff} is the following: while in \cite{Binoth:2005ff}
the numerical evaluation of box integrals was based on three-dimensional 
parameter representations, we use a certain one-dimensional parameter representation
here, obtained after performing two integrations analytically, as outlined in Appendix \ref{onedimint}.
In this way one can use deterministic integration routines, leading to a fast and precise numerical evaluation. This has been done for box integrals with up to three off-shell legs and all triangle 
integrals.
The relative error to be achieved in the numerical integration has been set to 
the default value $10^{-8}$. If this precision has not been reached, the program will 
write a message to the file {\tt error.txt}. 
In some cases, calculating in double precision Fortran may not be sufficient.
The code is designed such that  it can be compiled in quadruple precision as well. 

We would like to emphasize that the program also can be used as a library for 
master integrals with massless internal particles. For example, the scalar box integrals
in $n$ dimensions, with up to four off-shell external legs, can be calculated by 
just calling the form factor $A^{4,0}$. Depending on the kinematics, the 
program will call the appropriate box type automatically.
The scalar box integrals in $n+2$ and 
 $n+4$ dimensions are related to 
the form factors by $B^{4,2}=-I^{n+2}_4/2$, $C^{4,4}=I^{n+4}_4/4$, analogous for $N=3$.

\section{Overview of the software structure}

The structure of the {\tt golem95} program is the following:
There are four  main directories:
\begin{enumerate}
\item {\bf src:} the source files of the program
\item {\bf demos:} some programs for demonstration
\item {\bf doc:} documentation which has been created with robodoc~\cite{robodoc} 
\item {\bf test:} supplements the 
demonstration programs, containing files to produce form factors with user-defined kinematics. 
      The user can specify the rank, numerator, numerical point 
      etc. via a steering file.
\end{enumerate}

\section{Description of the individual software components}

Here we give a short summary of the contents of the individual modules.
A detailed description of the usage, dependencies and output of each 
module is given at the beginning of each file of the program
and can also be read in {\it html} format by loading 
the file {\tt masterindex.html} from the subdirectory {\tt doc}
into the browser and following the various links.\\
The program is written in Fortran95 and is downwards compatible 
to Fortran90. 

The directory {\bf src} contains the subdirectories 
\begin{itemize}
\item 
{\bf form\_factor:} contains five modules to compute the form factors 
for two-point to six-point functions:\\
          {\tt    form\_factor\_2p.f90, 
	     form\_factor\_3p.f90, 
	     form\_factor\_4p.f90},\\
           {\tt   form\_factor\_5p.f90, 
	     form\_factor\_6p.f90}.

\item {\bf integrals: } contains the subdirectories {\bf four\_point, three\_point, two\_point}.
\begin{description}
\item[four\_point:] contains six modules to compute the four-point functions
 with $p_i^2\not=0$ holding for four, three, two, one or none of the external legs:
            {\tt  function\_4p1m.f90, 
	     function\_4p2m\_opp.f90,
             function\_\-4p2m\-\_adj.f90, 
	     function\_4p3m.f90,
	     function\_4p4m.f90, 
	     generic\_function\_4p.\-f90}.

\item[three\_point:] contains six modules to compute the three-point functions
              with three, two or one external legs off-shell: \\
	     {\tt   function\_3p1m.f90, 
	      function\_3p2m.f90,
	      function\_3p3m.f90,}\\
	      {\tt  gene\-ric\_\-function\_3p.f90, 
	      mod\_h0.f90, mod\_hf.f90, mod\_he.f90}.

\item[two\_point:]  contains one module to compute the two-point functions: \\
             {\tt generic\_function\_2p.f90}.
\end{description}

\item {\bf kinematic:}  contains two modules to compute the matrix ${\cal S}$ 
and its inverse 
            and to compute the reduction coefficients $b_i$: \\
	   {\tt  matrice\_s.f90, inverse\_matrice.f90}.\\
The definition of $\hat{B}$ (see eq.~(\ref{bhat})\,) is contained in {\tt  matrice\_s.f90}.	   

\item {\bf module: } contains auxiliary functions/subroutines and the 
definition of some default parameters: \\
The file {\tt parametre.f90} contains the parameters defining the switch 
between the reduction 
down to scalar basis integrals (which are implemented in analytic form) 
versus the numerical evaluation of integrals 
(with or without Feynman parameters in the numerator), 
as explained in section \ref{basisints}. 
The default value for $\hat{B}$ for  three-point as well as four-point functions 
has been set to 0.005.
The other default parameters for the numerical integration
are also fixed in {\tt parametre.f90}.
Further, there is  a switch to calculate the rational parts of amplitudes 
only. The default is {\tt tot} to calculate the 
complete form factors. If  {\tt tot} is replaced by {\tt rat}, only the 
rational parts 
will be calculated.

\noindent
The auxiliary functions will not all be listed here, we only point to 
the most important ones:
\begin{itemize}
\item 
Polylogarithms and other special functions are defined in
{\tt z\_log.f90, zdilog.f90,  kronecker.f90, constante.f90}.
\item
{\tt spinor.f90} contains  functions to compute scalar products of four-momenta,
spinorial products and totally antisymmetric epsilon tensors.
\item
The files {\tt preci\_double.f90} and 
{\tt preci\_quad.\-f90} are needed for the switch between double precision and 
quadruple precision. The default is double precision.
If quadruple precision should be used, 
one has to define {\tt \$precision = "quadruple"} in the file {\tt configure.pl}.
Note that quadruple precision is at present only supported by the {\tt ifort} compiler.
\item
The file {\tt form\_factor\_type.f90} defines a type {\it form\_factor} 
such that form factors, which are arrays of three complex numbers, 
can be involved in algebraic manipulations.
\item
{\tt cache.f90}  is used to reserve  memory in order to store results for three-point or 
four-point functions which already have been computed.
\end{itemize}	   
\item {\bf numerical: } contains two modules for the numerical integration : \\
	  {\tt mod\_adapt\_\-gauss.\-f90,  mod\_numeric.f90}.
\end{itemize}
Concerning the numerical integration, the following features should be pointed out:
\begin{itemize}
\item The user can change the integration method for the numerical integration
of the one-dimensional parameter integrals by 
changing the module {\tt numerical\_evaluation} in the file {\tt mod\_numeric.f90 }
in the directory \\
{\tt src/nu\-me\-ri\-cal}.
\item The values for the cuts defining the switch to a one-dimensional numerical
integration of the basis integrals are given in {\tt parametre.f90}
and can be changed easily by the user.\\
{\em Note:} If the user wants to change the 
default values defined in {\tt parametre.\-f90}, it is 
{\em not} necessary to recompile the library. 
If the desired values are defined in the main program, 
the default values will be overwritten. 
The command {\tt use parametre} still  has to be included in the header of the main program. 
\item  For boxes with 4 off-shell external legs, the expressions 
for one-dimensional numerical integrations are not worked out 
in this version. Here the program will always reduce numerically to scalar 
basis integrals, irrespective of the size of the Gram  determinants.
\end{itemize}

\section{Installation instructions}

The program can be downloaded as a  .tar.gz archive  from the following URL:
{\tt http://lappweb.in2p3.fr/lapth/Golem/golem95\_v1.0.tar.gz}. 
The installation instructions given below also can be found in the {\tt Readme} file 
coming with the code.

To install the {\tt golem95} library, type the following commands:\\
{\tt ./configure.pl [--install\_path=mypath] [--compiler=mycompiler]}\\
{\tt make}\\
{\tt make install}

Please note that {\tt mypath} must be the absolute path of the directory 
where you would like the library to be intalled. 
If no option for {\tt install\_path} is given, 
a subdirectory  of  
the current directory  with the name {\tt libgolem} 
will be created and the library will be installed in 
this subdirectory.

For example, if you want to put the library into the directory \\
{\tt /home/myname/lib/libgolem} and use the compiler {\tt g95}, then type:\\
{\tt ./configure.pl --install\_path=/home/myname/lib/libgolem --compi\-ler=g95}\\
{\tt make}\\
{\tt make install}

The directory {\tt /home/myname/lib/libgolem} will then contain a collection of 
files of type {\tt .mod} plus a file named {\tt libgolem.a} which is the {\tt golem95} library. 

If no option for the compiler is specified, the installation script will search for 
fortran 95 compilers installed on your system and will take the first 
matching compiler found.

The program has been tested with the GNU compilers g95 and gfortran, 
the intel compiler ifort,  
the dec compiler f95, the NAG compiler f95 and 
the portland compiler pgf95.

\section{Test run description}

The program comes with several demonstration programs located in the subdirectory 
{\tt demos}.  
A list of all options  contained in the {\tt demos} directory is given in Appendix \ref{demos}.
We will describe some selected examples in the following. 

\subsection{Rank five five-point form factor}
As an example for a test run, we first describe the calculation of  
a form factor for a rank five 5-point integral, $A^{5,5}_{j_1\ldots j_5}$. 
We choose $j_i=i$, i.e. $z_1\ldots z_5$ in the numerator.
Further, we choose the following numerical point (in terms of entries 
of the kinematic matrix ${\cal S}$ containing the invariants):
\begin{equation}
{\cal S} =
\left( \begin{array}{ccccc} 
0&p_2^2&s_{23}&s_{51}&p_1^2\\
p_2^2&0&p_3^2&s_{34}&s_{12}\\
s_{23}&p_3^2&0&p_4^2&s_{45}\\
s_{51}&s_{34}&p_4^2&0&p_5^2\\
p_1^2&s_{12}&s_{45}&p_5^2&0
\end{array}\right)
=\left( \begin{array}{ccccc} 
0&0&-3&-4&0\\
0&0&0&6&15\\
-3&0&0&0&2\\
-4&6&0&0&0\\
0&15&2&0&0
\end{array}\right)
\end{equation}
These values are already implemented in the file {\tt demo\_5point.f90} 
in the subdirectory {\tt demos}.
All the user has to do is the following:
\begin{itemize}
\item go to the subdirectory {\tt demos}
\item type``perl configure.pl". The shell will prompt for the choice of 
the demo to be run:\\
{\tt Choose which demo program you want to run:\\
1) three-point functions\\
2) four-point functions\\
3) five-point functions\\
4) six-point functions\\
5) 4-photon helicity amplitudes\\
6) numerical stability demo: $\det G\to 0$\\
7) numerical stability demo: $\det S\to 0$\\
8) Golem $\leftrightarrow$ LoopTools conventions
}
\item 
Choosing  option 3 will produce the following output:\\
{\tt 
you have chosen option 3: five-point functions\\
The Makefile has been created\\
Please run:\\
make\\
./comp.exe\\
}
\item Running ``make" will produce the executable {\tt comp.exe} where the 
{\tt demo*.f90} files matching the choice above will be compiled automatically.
Running {\tt comp.exe} will prompt for the rank of the form factor to be calculated:\\
{\tt 
Choose what the program should compute:\\
 0) form factor for five-point function, rank 0\\
 1) form factor for five-point function, rank 3 (z1*z2*z4)\\
 2) form factor for five-point function, rank 5 (z1*z2*z3*z4*z5)\\
 3) form factor for diagram with propagator 3 pinched, rank 0\\
 4) form factor for diagram with propagators 1 and 4 pinched, rank0
}
\item Choosing option 2 will produce the result 
which will be written to the file {\tt test5point.txt} and looks as follows:


 The kinematics is:
\begin{picture}(150,60)
\put(135,-150){\includegraphics[width=4cm, height=4cm]{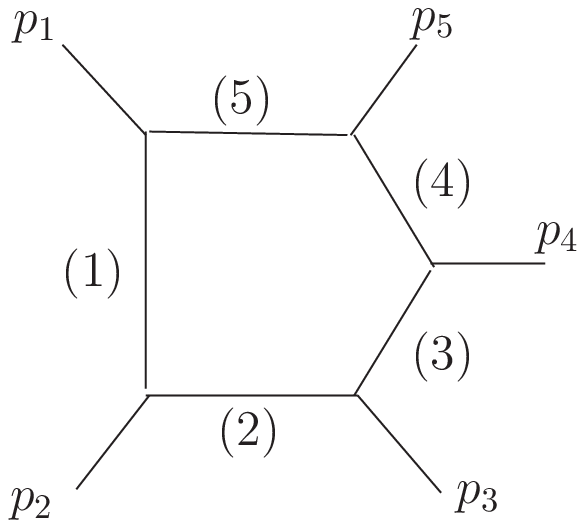}}
\end{picture}
\begin{eqnarray*} 
&& p_1+p_2+p_3+p_4+p_5 = 0\\
  &&S(1,3) = (p_2+p_3)^2=-3.\\
  &&S(2,4) = (p_3+p_4)^2=6.\\
  &&S(2,5) = (p_1+p_2)^2=15.\\
  &&S(3,5) = (p_4+p_5)^2=2.\\
  &&S(1,4) = (p_1+p_5)^2=-4.\\
  &&S(1,2) = p_2^2=0.\\
  &&S(2,3) = p_3^2=0.\\
  &&S(3,4) = p_4^2=0.\\
  &&S(4,5) = p_5^2=0.\\
  &&S(1,5) = p_1^2=0.
\end{eqnarray*} 
A factor $\Gamma(1+\epsilon) \Gamma(1-\epsilon)^2/\Gamma(1-2 \epsilon)\,(4\pi\,\mu^2)^{\epsilon}$
 is factored out from  the result.\\
\begin{tabular}{ll} 
result=& 1/$\epsilon^2$ * (0.0000000000E+00 +  I* 0.0000000000E+00)\\
&+ 1/$\epsilon$ * (0.0000000000E+00 +  I* 0.0000000000E+00)\\
&+ (--.8615520644E-04 +  I* 0.1230709464E-03)\\
CPU time=&  7.999000000000001E-003
\end{tabular}
\end{itemize}
We recall that  we use the integral measure as in eq.~(\ref{eq0}).
The factor 
$r_\Gamma=\Gamma(1+\epsilon) \Gamma(1-\epsilon)^2/\Gamma(1-2 \epsilon)\,(4\pi\,\mu^2)^{\epsilon}$
has been extracted from the integrals to comply with the conventions 
of ref.~\cite{Binoth:2005ff} and the 
$\overline{\rm{MS}}$ subtraction scheme. 
Note that it may be advantageous to call {\tt golem95} with rescaled, dimensionless  
invariants (e.g. $s_{ij}/\mu^2$), for example in cases where most of the invariants 
have very small numerical values in all kinematic regions. 

\subsection{Calculating all possible numerators at once}

In the previous example, the  numerical point (and the type of numerator 
for tensor integrals) has been fixed in the demo programs. 
If the user would like to give the the numerical point and the 
Feynman parameters in the numerator as an {\it input}, he 
can use the file {\tt param.input} in the subdirectory {\tt test}.
This setup also allows to calculate all possible numerators for a certain rank 
in one go.
A typical example  looks as follows:

Assume we would like to calculate the form factors for rank one six-point functions
for all possible numerators $z_j, j=1\ldots 6$, for the following numerical point
($p_i=(E_i,x_i,y_i,z_i)$):
\bea
p_1&=&(0.5,0.,0., 0.5)\nn\\
p_2&=&(0.5,0.,0.,-0.5)\nn\\
p_3&=&(-0.19178191 ,-0.12741180,-0.08262477,-0.11713105)\nn\\
p_4&=&(-0.33662712, 0.06648281, 0.31893785, 0.08471424)\nn\\
p_5&=&(-0.21604814, 0.20363139,-0.04415762, -0.05710657)\nn\\
p_6&=&-\sum_{i=1}^5 p_i \nn
\eea
To use these momenta, go to the subdirectory {\tt test} and 
edit\footnote{Alternatively, random momenta can be generated using the 
program mom\_rambo.f, adapted from \cite{Kleiss:1985gy} and 
also contained in the subdirectory {\tt test}.}
 the file 
{\tt momenta.\-dat}, writing each component of the above momenta  
into a single line.
To calculate an $N$-point function,  
   the program will use the first $N$ momenta found in 
   {\tt momenta.dat} (respectively the momenta file specified in {\tt param.input}).
   For $N=5$ and $N=6$, it is important that momentum 
   conservation, i.e. $\sum_{i=1}^N p_i = 0$, is fulfilled because 
   momentum conservation has been assumed to hold true in the reduction.
   
To generate results, the user only has to do the following:
\begin{itemize}   
 \item  edit the file {\tt param.input} to choose the number of legs, rank and numerator.
If  only a particular numerator should be calculated, 
 give the labels of Feynman parameters, else put {\tt all} into the numerator field, 
 \item type {\tt perl maketest.pl}. 
 \end{itemize}
 The program will automatically compile the corresponding functions 
 and run the  executable. 
  The following output will be produced:
\begin{enumerate}
\item  a separate output file called N[nb of legs][rank]\_[pt].out
     for each individual numerator
     ([pt] denotes the ``label" of a particular numerical point, 
       which can be chosen by the user to distinguish results
       for different numerical points)
\item a file called N[nb of legs][rank]\_[pt].numbers,
       where all form factors that have been calculated 
       for a particular rank and number of legs and numerical point 
       are {\it appended}. 
       For example, if the option {\tt all} has been chosen to 
       calculate all possible combinations of Feynman parameters 
       in the numerator, this file will contain the results for all 
       these numerators.
       The format is such that it can be read by {\tt Mathematica},  
       to allow direct comparisons to results obtained from algebraic programs.
       If the result is $P_2/\eps^2+P_1/\eps+P_0$ for a 
       rank $r$ $N$-point form factor of type $A$, 
       the output will be a list
       $a_N[j_1,\ldots,j_r]=\{{\cal R}e[P_2],{\cal I}m[P_2],{\cal R}e[P_1],{\cal I}m[P_1],{\cal R}e[P_0],{\cal I}m[P_0]\}$.
       \end{enumerate}
       For example, for rank one six-point functions at the numerical 
       point given above (``pt1"), having chosen {\tt all} in {\tt param.input} 
       to calculate all six possible 
       numerators, the program produces seven output files:
       {\tt N6rank1zi\_pt1.out} for $i=1\ldots 6$ and the file
       {\tt N6rank1\_pt1.numbers}. 
       While the files {\tt N6rank1zi\_pt1.out} contain, in addition to the result 
       for the particular numerator, also the kinematic point and CPU time information, 
       the file  {\tt N6rank1\_pt1.numbers} just lists the results.
Note that for $N\ge 5$,  individual form factors 
are not uniquely   defined because the metric tensor   
$g_{\mu\nu}$ can be expressed by external 
momenta, such that individual terms can be shifted between 
form factors of type $A,B$ or $C$.   
   
\subsection{Calculation of the 4-photon helicity amplitudes}

In order to show how {\tt golem95} can be embedded into the 
calculation of full one-loop amplitudes, we give here the 
calculation of the light-by-light scattering  amplitude in massless QED
as a pedagogical example.

This amplitude is defined by
six Feynman diagrams where the four photons are attached to a closed
fermion loop in all possible ways. Diagrams 
which differ by the charge flow only lead to the same value,
which leaves us with three different topologies defined by the
photon orderings $1243$, $1234$ and $1324$, respectively.  
Each diagram is IR finite and UV divergent. The UV divergence only
cancels in the sum of the diagrams. 
The results for the three independent helicity amplitudes
$++++$, $+++-$, $++--$ are well known, see for example 
\cite{Binoth:2002xg,Bernicot:2008th}. 
For completeness we list the analytic formulae, omitting
the irrelevant phases
\begin{eqnarray}
{\cal A}^{++++} &=& 8 \quad , \quad 
{\cal A}^{+++-} = -8 \;,\nn \\
{\cal A}^{++--} &=& -8 \Bigl[  1 + \frac{t-u}{s} \log\left(\frac{t}{u}\right) 
 + \frac{t^2+u^2}{2 s^2} \Bigl( \log\left(\frac{t}{u}\right)^2 + \pi^2 \Bigr)\Bigr] \;\; .
\end{eqnarray}

The analytic expressions in terms of 
form factors and Mandelstam invariants 
which are given in the demo program {\tt demo\_4photon.f90} were obtained as
follows. 
After working out the trace of gamma matrices one finds
for each graph a polynomial in scalar products of polarization
vectors, $\varepsilon_j$, external momenta $p_j$ 
and the $D=4-2\epsilon$ dimensional loop momentum $k$.
All reducible scalar products, i.e. those which can be written in terms of inverse propagators,
were cancelled directly. The remaining expressions, containing only
irreducible scalar products, are proportional to tensor integrals
which  are transformed to  form factors using eq.~(\ref{fofageneral}).
Each form factor now has scalar coefficients containing 
polarisation vectors and external momenta, i.e.
$\varepsilon_i \cdot \varepsilon_j$, $\varepsilon_i \cdot p_j$, 
$s=2 p_1 \cdot p_2$, $t=2 p_2\cdot p_3$ and $u=2 p_1\cdot p_3$, 
where we defined all external momenta as incoming.
Using spinor helicity methods 
one can map these coefficients to polynomials in the Mandelstam variables $s$, $t$ and $u=-s-t$.
Choosing reference momenta $p_2$, $p_1$, $p_4$,
$p_3$ for the polarization vectors $\varepsilon_1$,   $\varepsilon_2$,   
$\varepsilon_3$, $\varepsilon_4$ respectively,  
one easily can show the following relations~\cite{Binoth:2003xk}
relevant for the $++++$ amplitude
\begin{eqnarray}
\varepsilon_1^+ \cdot \varepsilon_1^+ = -\frac{2 s}{tu} \varepsilon_1^+ \cdot p_3\,\varepsilon_2^+ \cdot p_3 &,&
\varepsilon_1^+ \cdot \varepsilon_3^+ =  \frac{2  }{t } \varepsilon_1^+ \cdot p_4\,\varepsilon_3^+ \cdot p_1 \;,\nn\\
\varepsilon_1^+ \cdot \varepsilon_4^+ =  \frac{2  }{u } \varepsilon_1^+ \cdot p_3\,\varepsilon_4^+ \cdot p_1 &,&
\varepsilon_2^+ \cdot \varepsilon_3^+ =  \frac{2 s}{u } \varepsilon_2^+ \cdot p_4\,\varepsilon_3^+ \cdot p_2 \;,\nn\\
\varepsilon_2^+ \cdot \varepsilon_4^+ =  \frac{2 s}{t } \varepsilon_2^+ \cdot p_3\,\varepsilon_4^+ \cdot p_2 & ,&
\varepsilon_3^+ \cdot \varepsilon_4^+ = -\frac{2 s}{tu} \varepsilon_3^+ \cdot p_1\,\varepsilon_4^+ \cdot p_1 \;,\nn\\
\varepsilon_1^+ \cdot p_3\,\varepsilon_2^+ \cdot p_3 \,\varepsilon_3^+ \cdot p_1\, \varepsilon_4^+ \cdot p_1
 &=& \left(\frac{tu}{2s}\right)^2 \frac{[21][43]}{\langle 12 \rangle  \langle 34\rangle}\;.
 \label{4p}
\end{eqnarray}  
The phase factor in the last line is irrelevant for observables and thus can be dropped.
All coefficients of the form factors of the $++++$ amplitude are now rational polynomials in $s$, $t$ and $u=-s-t$.
For the $+++-$ amplitude one needs instead of eq.\,(\ref{4p})
\begin{eqnarray}
\varepsilon_1^+ \cdot \varepsilon_4^- =  \frac{2  }{t } \varepsilon_1^+ \cdot p_4\,\varepsilon_4^+ \cdot p_1 &,&
\varepsilon_2^+ \cdot \varepsilon_4^- =  \frac{2  }{u } \varepsilon_2^+ \cdot p_4\,\varepsilon_4^+ \cdot p_2 \;,\nn\\
\varepsilon_3^+ \cdot \varepsilon_4^- =  0\qquad\qquad\qquad \;,\nn\\
\varepsilon_1^+ \cdot p_3\,\varepsilon_2^+ \cdot p_3 \,\varepsilon_3^+ \cdot p_1\,\varepsilon_4^- \cdot p_1
 &=& \left(\frac{tu}{2s}\right)^2 \frac{[21] \langle 14 \rangle [31]}{\langle 12 \rangle [41]  \langle 13 \rangle}
\end{eqnarray}  
and for the $++--$ amplitude
\begin{eqnarray}
\varepsilon_1^+ \cdot \varepsilon_3^- =  \frac{2  }{u } \varepsilon_1^+ \cdot p_3\,\varepsilon_3^+ \cdot p_1 &,&
\varepsilon_1^+ \cdot \varepsilon_4^- =  \frac{2  }{t } \varepsilon_1^+ \cdot p_4\,\varepsilon_4^+ \cdot p_1 \;,\nn\\
\varepsilon_2^+ \cdot \varepsilon_3^- =  \frac{2  }{t } \varepsilon_2^+ \cdot p_3\,\varepsilon_3^+ \cdot p_2 &,&
\varepsilon_3^- \cdot \varepsilon_4^- = -\frac{2 s}{tu} \varepsilon_3^- \cdot p_1\,\varepsilon_4^- \cdot p_1  \;,\nn\\
\varepsilon_1^+ \cdot p_3\,\varepsilon_2^+ \cdot p_3 \,\varepsilon_3^- \cdot p_1\,\varepsilon_4^- \cdot p_1
 &=& \left(\frac{tu}{2s}\right)^2 \frac{[21]\langle 34 \rangle }{\langle 12 \rangle [ 34 ]}\;.
\end{eqnarray}  
These relations define the different coefficients of the 
form factors present in the file {\tt demo\_4photon.f90} which evaluates the 
four photon amplitude\footnote{We note that the three helicity amplitudes can be evaluated in a much simpler way.
By applying spinor helicity methods at an earlier stage, one can achieve a representation without
any tensor four-point function. The given representation should only illustrate
a generic form factor representation of  an amplitude.}.
The form 
factors for each momentum ordering have to be evaluated only once. 
We have compared our numerical result  with the well-known results for these amplitudes
and find perfect agreement. 

The program {\tt demo\_4photon.f90} can be used as a guideline how 
to express any amplitude with massless 
loops  in terms of form factors and scalar coefficients before evaluating it 
with {\tt golem95}.

\section{Conclusions and Outlook}

We have presented the Fortran 95 program {\tt golem95} for the numerical evaluation of 
tensor integrals  up to rank six six-point functions.
The program is based on a form factor representation of tensor integrals and
performs the reduction to a certain set of basis integrals numerically.
The basis integrals are implemented in analytic form. 
If during the reduction process an inverse determinant becomes small, the program 
switches to a numerical evaluation of  the (tensor-)integral without further 
reduction, thus avoiding small denominators. 
The numerical evaluation is based on one-dimensional parameter integral representations 
for most of the basis integrals, allowing for a fast and precise numerical integration.

The results are given as
a set of three complex numbers representing the 
coefficients of the Laurent series in the dimensional regularisation parameter $\epsilon$, 
i.e. the coefficients of the $1/\epsilon^2, 1/\epsilon$ poles and the finite part.

The program can also be used as a library for master integrals (including infrared 
divergent ones), as the form factors with no Feynman parameter labels 
directly correspond to scalar integrals.
In the current version, master integrals with massive {\it internal} particles 
are not implemented yet. They will be available in a forthcoming version.
There is no restriction on the number of massive {\it external } legs.

A future version will also combine the {\tt golem95} code for the form factor evaluation 
with a code for the generation of  amplitudes, thus moving towards a 
full automatisation of the calculation of one-loop amplitudes.

\section*{Acknowledgements}
We would like to thank A.~Guffanti and G.~Sanguinetti for collaboration at an earlier stage of this work. 
TB, GH and TR would like to thank the LAPTH for hospitality 
while part of this work was carried out.
This research was supported by the UK Science and Technology Facilities Council 
(STFC) and the Scottish Universities Physics Alliance (SUPA).

\renewcommand \thesection{\Alph{section}}
\setcounter{section}{0}

\section{Appendices}
\renewcommand{\theequation}{\Alph{section}.\arabic{equation}}
\setcounter{equation}{0}


\subsection{Landau singularities}\label{landau}

Besides the spurious appearence of powers of inverse Gram determinants caused 
by the decomposition on scalar integrals in the reduction process, which can be
avoided e.g. with the  method advocated here and in \cite{Binoth:2005ff}, 
another source of problems in the numerical
evaluation of scattering amplitudes may be caused by the occurence of 
{\em actual} kinematic singularities, the so-called Landau singularities 
\cite{ELOP}. 
The latter may appear in some diagrams contributing to the considered amplitude 
whenever the determinant  of the kinematic matrix ${\cal S}$ associated with 
these diagrams - or with reduced diagrams obtained by one or several pinches 
- vanishes. Typical cases of such singularities are 
threshold singularities in loop calculations with internal and external masses, 
collinear and infra-red singularities with massless internal and external lines. 
Another type is the one of scattering singularities 
\cite{Bern:2008ef,Nagy:2006xy}, for which $(\det G)\to 0$ and
$\det S$ becomes proportional to $(\det G)^{2}$, 
such that both
 vanish simultaneously. The occurence of these particular  cases of  
vanishing Gram determinants should not be confused with the spurious ones. 
Individual 
diagrams lead to infinities at such kinematic configurations, where a mass
singularity and a scattering singularity coincide.

In addition, it should be noted that for a given diagram, the reduction 
algorithm breaks down\footnote{In the example of double parton scattering for $2$(massless) 
$\to 2$ massive legs with no internal mass, it can be checked that for the four 
leg diagram with two opposite masses, the equations determining $B$ and $b_{4}$ 
in the notations of ref. \cite{Binoth:2005ff} have no solution, 
because $(\delta v).H.(\delta v) = 0$, hence the equation for $B$ becomes 
$0\times B=1$ which has obviously no solution. For a discussion of 
the reduction in exceptional kinematic configurations 
see also \cite{Duplancic:2003tv}.} at such a scattering singularity,
as inferred from the relation $B \propto \det G / \det S \to \infty$. 
As one combines the diagrams to scattering amplitudes, 
gauge  cancellations may occur analytically, which in general
reduces the degree of singularity as compared to  individual 
diagrams, or even make the singularity bounded, as e.g. observed in 
the 6 photon amplitudes\footnote{{\em Singularity} means {\em non-analyticity}; 
the latter can be either infinite - integrable or 
not - or bounded.} \cite{Bern:2008ef,Bernicot:2007hs}. 
On the other hand, the numerical combination 
of the singularities from separate diagrams is expected to be problematic, and 
leads to instabilities even in cases of expected finiteness. Note that this 
problem is known to all methods based on the reduction of diagrams, so it is
is not specific to our reduction formalism. 
We note that the problem  of large numerical cancellations 
is inherent to any method based on the reduction to scalar master integrals
like $I_3^n, I_4^n$, as the latter may become linearly dependent near such singularities.

Depending on the inclusiveness of the observable to be calculated, 
 and the degree of the singularity, possible cures could be to 
resort to mutliple precision in some vicinity of the kinematic singularities, 
and/or place a hole in the phase space around the singularity together with a smooth
interpolation over it. Certainly, in specific cases,  
when the observable would be controlled 
by the singularity, such  methods would be inadequate.   

\subsection{One-dimensional integral representations}\label{onedimint}

In this appendix we will derive representations of 
IR finite box-and triangle integrals as one-dimensional 
Feynman parameter integrals.
These representations have the advantage that they can be integrated 
numerically in a very fast and precise way using deterministic 
numerical integration routines.  This approach is similar to the one in ~\cite{Binoth:2002xh}
where one parameter integration has been carried out analytically.
The program switches to this numerical evaluation if 
$\hat{B}$ becomes smaller than a value defined in {\tt module/parametre.f90}
(the default is 0.005).

\subsubsection{Four-point functions}\label{secOMFPF}

Our starting point are the higher dimensional four-point functions $I^{n+2}_4,I^{n+4}_4$
given by
\begin{eqnarray}
I^{n+2}_4(j_1, \ldots ,j_r) &=& 
\Gamma \left(3-\frac{n}{2} \right) \, \int_{0}^{1} 
\prod_{i=1}^{4} \, d z_i \, \delta(1-\sum_{l=1}^{4} z_l) 
\, \frac{z_{j_1} \ldots z_{j_r}}{ 
(-\frac{1}{2}\, z \cdot \calst
\cdot z-i\delta)^{3-n/2}}\;,\nn\\
I^{n+4}_4(j_1) &=& 
\Gamma \left(2-\frac{n}{2} \right) \, \int_{0}^{1} 
\prod_{i=1}^{4} \, d z_i \, \delta(1-\sum_{l=1}^{4} z_l) 
\, \frac{z_{j_1}}{ 
(-\frac{1}{2}\, z \cdot \calst
\cdot z-i\delta)^{2-n/2}}\;,\nn
\end{eqnarray}
The reduction of these integrals to integrals with no Feynman parameters in 
the numerator introduces inverse Gram determinants.
Therefore it can be advantageous to evaluate these integrals without further 
reduction. To do so, we proceed as follows:\\
First, to get rid of the $\delta$ distribution, we make the following change of variables:
\begin{eqnarray}
z_1 & = & w \, (1-x)\; ,\;
z_2  =  w \, x \, y \, z \; ,\;
z_3  =  w \, x \, y \, (1-z) \; ,\;
z_4  =  w \, x \, (1-y) 
\label{eqCHANGEV}
\end{eqnarray}
Now,  instead of computing directly the three-dimensional integrals numerically as 
proposed in \cite{Binoth:2005ff}, we perform analytically the integration over $x$ 
and $y$ and integrate numerically over the leftover variable $z$, using  
an adaptive Gauss-Kronrod method~\cite{numrecipes}.

For the cases treated in the {\tt golem95} library (no internal masses), the $x$ and $y$
 integration for the six- and eight-dimensional four-point functions can be computed 
 using two basis integrals:
\bea
\int^1_0 \, dx \, \frac{x^n}{A + B \, x} &=& J(n,A,B)\;,
\label{eqCAS1}\\
\int^1_0 \, dx \, x^n \, \ln(A + B \, x) &=& K(n,A,B)
\label{eqCAS2}
\eea
which obey to the following relations:
\begin{eqnarray}
J(n,A,B) & = & \frac{1}{n \,B} - \frac{A}{B} \, J(n-1,A,B) \\
J(0,A,B) & = & \frac{\ln(A+B) - \ln(A) }{B} \\
K(n,A,B) & = & \frac{ (A+B) \, \ln(A+B) - n \, A \, K(n-1,A,B) }{(n+1) \, B} \nn \\
& & \mbox{} - \frac{1}{(n+1)^2} \\
K(0,A,B) & = & \frac{ (A+B) \, \ln(A+B) - A \, \ln(A) -B }{B} 
\label{eqPROP}
\end{eqnarray}
Here we assume that $B \ne 0$; if $B=0$ the integrations are trivial.
When the two first integrations have been done, we are left with the $z$ integration. 

To explain how we proceed, we treat the case of the six-dimensional 
three-mass four-point function as an example. 
After integration over $x$ and $y$, we are left with the following structure:
\begin{eqnarray}
I & = &  \frac{ - h \, \ln(h) + e \, \ln(e) }{f \, g}  + 
\frac{ h \, \ln(h)  - c \, \ln(c) }{f \, d}
\label{eqZDEPEND}
\end{eqnarray}
with
\begin{eqnarray}
c & = & z \, \calst_{12}+(1-z) \, \calst_{13}\nn \\
f & = & z \, (\calst_{24}-\calst_{12})+(1-z) \, (\calst_{34}-\calst_{13})\nn \\
g & = & z \, (1-z) \, \calst_{23}-z \, \calst_{24}-(1-z) \, \calst_{34}\nn \\
d & = & z \, (1-z) \, \calst_{23}-z \, \calst_{12}-(1-z) \, \calst_{13}\nn \\
e & = & z \, \calst_{24}+(1-z) \, \calst_{34}\nn \\
h & = & z \, (1-z) \, \calst_{23} 
\label{eq DEFCFGDEH}
\end{eqnarray}
where $\calst_{ij}$ are the $\calst$-matrix elements, they must be understood 
as $\calst_{ij} + i \, \delta$ .

The first thing to note is that $I$ has no poles. All six-dimensional 
four-point functions are infrared finite, and the UV pole of the 
eight-dimensional four-point functions is contained in the overall 
$\Gamma$-function.
Indeed, it is easy to see that: $g=h-e$, $d=h-c$ and $f=e-c$, so when 
$g \rightarrow 0$ or $d \rightarrow 0$ or $f \rightarrow 0$, the numerator 
of $I$ goes to zero.  
To compute the $z$ integral numerically, we use a contour deformation: 
we complexify the $z$ variable
\begin{equation}
z = u - i \, \epsilon \, g(u)
\label{eqCHANGEZ}
\end{equation}
i.e. we have to compute the following integrals:
\begin{equation}
\int^1_0 \, dz \, f(z) = \int^1_0 \, du \, C \, f(u - i \, \epsilon \, g(u))
\label{eqTRANSZU}
\end{equation}
where $C$ is the jacobian of the transformation : 
$C = 1 - i \, \epsilon \, dg/du$ and $\epsilon = \pm1$. 
The function $g$ has the following properties: $g(0) = g(1) = 0$ and $g(u) > 0$ for $u \in [0,1]$. 
For practical applications, we took $g(u) = u \, (1-u)$.
As the numerator of $I$ contains some logarithms,  some care has to be taken in order to avoid a clash between the cut of the logarithm and the contour. To analyse that, let us consider the following example:
\begin{equation}
E = \int^1_0 \, dz \, \frac{\ln(a+b \, z + i \, s_1 \, \lambda)}{c + d \, z + i \, s_2 \, \lambda}
\label{eqEXAMPLEE}
\end{equation}
with $a$, $b$, $c$ and $d \in \mathbb{R}$, $s_1, \, s_2 = \pm1$ and $\lambda > 0$. Making the change of variable (\ref{eqCHANGEZ}) , we get:
\begin{equation}
E = \int^1_0 \, du \, C \, \frac{\ln(a+b \, u + i \, (s_1 \, \lambda - b \, \epsilon g(u)) )}{c + d \, u  + i \, (s_2 \, \lambda - d \, \epsilon \, g(u)) }
\label{eqEXAMPLEE1}
\end{equation}
By choosing $\epsilon = - \, s_1 \, \mbox{sign}(b)$, the imaginary part of the argument of the
 logarithm  will be constant and have the sign of $s_1$. This choice of $\epsilon$ defines 
 the contour but the important point is that by varying $u$ (walking on the contour) the cut 
 of the logarithm is never crossed. The pole is located at:
\begin{equation}
z_0 = - \frac{c}{d} - i \, \frac{s_2}{d} \, \lambda
\label{eqLOCATIONOFTHEPOLE}
\end{equation}
Using the Cauchy theorem, we arrive at the following relation:
\begin{eqnarray}
\int^1_0 \, dz \, f(z) & = & \int^1_0 \, du \, C \, \frac{\ln(a+b \, u + i \, (s_1 \, \lambda - b \, \epsilon g(u)) )}{c + d \, u  + i \, (s_2 \, \lambda - d \, \epsilon \, g(u)) } \nn \\
& & \mbox{} - 2 \, i \, \pi \, R \, \epsilon \, \Theta \left(1+\frac{c}{d}\right) \, \Theta \left(-\frac{c}{d}\right) \, \delta^{\rm{sign}(\epsilon)}_{\rm{sign}(s_2/d)}
\label{eqRESFINEX}
\end{eqnarray}
where $R$ is the residue of $f(z)$ at $z = z_0$.
This is the way we proceed to compute numerically the two terms of eq. (\ref{eqZDEPEND}). We compute the two terms separately despite the fact that each term has a pole when $f \rightarrow 0$ 
while the sum does not, because they contain two kinds of logarithms ($\ln(e)$ and $\ln(c)$),
 and there is no reason that the choice for $\epsilon$ for one term prevents the contour 
 from crossing the cut for the other term.

For the case where there are Feynman parameters in the numerator, everything works 
like the preceding example: we always split the integrand of the $z$ integration into 
two pieces (each piece having more terms than the scalar case) by separating the two 
kinds of logarithms. For the other types of four-point functions, we proceed in an
analogous way.

\subsubsection{Three-mass three-point functions}\label{secT3MTPF}

In the case of the  three-point functions with three off-shell legs, 
after making a change of variables of type (\ref{eqCHANGEV}), 
we are left with two-dimensional integrals. 
One parameter is integrated out analytically using (\ref{eqCAS1}), 
the remaining integral is computed numerically, using the same techniques as
 for the four-point functions.

\subsubsection{Two-mass three-point functions}\label{secTMTPF}
The two mass three-point functions are written in terms of functions $H_i$ \cite{Binoth:2005ff}, 
which are defined such that in the numerically problematic case where $X\to Y$, 
their evaluation is numerically stable.
The  functions $H_0$, $H_1$, $H_2$, $H_3$ and $H_4$ are given by:
\begin{eqnarray}
H_0(X,\alpha) & = & \frac{\bar{X}^{\alpha}}{X} \\
\label{eqH0}
H_1(X,Y,\alpha) & = & \frac{\bar{X}^{\alpha}-\bar{Y}^{\alpha}}{X-Y} \\
\label{eqH1}
H_2(X,Y,\alpha) & = & \frac{\bar{Y}^{\alpha}}{Y-X}+\frac{1}{1+\alpha} \, 
\frac{\bar{Y}^{1+\alpha}-\bar{X}^{1+\alpha}}{(Y-X)^2} \\
\label{eqH2}
H_3(X,Y,\alpha) & = & \frac{\bar{Y}^{\alpha}}{Y-X}+\frac{2}{1+\alpha} \, 
\frac{\bar{Y}^{1+\alpha}}{(Y-X)^2} \nn \\
& & \mbox{}+\frac{2}{(1+\alpha) \, (2+\alpha)} \, 
\frac{\bar{Y}^{2+\alpha}-\bar{X}^{2+\alpha}}{(Y-X)^3} \\
\label{eqH3}
H_4(X,Y,\alpha) & = & \frac{\bar{Y}^{\alpha}}{Y-X}+\frac{3}{1+\alpha} \, 
\frac{\bar{Y}^{1+\alpha}}{(Y-X)^2}+\frac{6}{(1+\alpha) \, (2+\alpha)} \, 
\frac{\bar{Y}^{2+\alpha}}{(Y-X)^3} \nn \\
& & \mbox{}+\frac{6}{(1+\alpha) \, (2+\alpha) \, (3+\alpha)} \, 
\frac{\bar{Y}^{3+\alpha}-\bar{X}^{3+\alpha}}{(Y-X)^4}\label{eqH4}\\
\bar{X}&=&-X-i\,\delta\nn
\end{eqnarray}
For each function $H_i(X,Y,\epsilon)$, one can define
\begin{equation}
H_i(X,Y,\epsilon) = \epsilon \, H_{E_i}(X,Y) + \frac{\epsilon^2}{2} \, H_{F_i}(X,Y)\;,
\end{equation}
and one can show that
\begin{eqnarray}
H_{E_n}(X,Y) & = & \int^1_0 dz \, z^{(n-1)} \; \frac{1}{z \, \bar{X}+(1-z) \, \bar{Y}} \label{eqDEFNMHE}\\
H_{F_n}(X,Y) & = & \int^1_0 dz \, z^{(n-1)} \; \frac{\ln(z \, \bar{X}+(1-z) \, \bar{Y})}{z \, \bar{X}+(1-z) \, \bar{Y}}
 \label{eqDEFNMHF}
\end{eqnarray}
From this definition, it is easy to show that
\begin{eqnarray}
H_{E_n}(X,Y) & = &  \frac{1}{X-Y} \, \left( \frac{1}{n-1} -Y \,H_{E_{n-1}}(X,Y) \right)
\label{eqHEREC}
\end{eqnarray}
The equations (\ref{eqDEFNMHE}) and (\ref{eqDEFNMHF}) are used to compute numerically the functions $H_{E_n}$ and $H_{F_n}$.

\subsection{Contents of the demonstration programs}\label{demos}

The demo programs calculate the following examples, listed also in the file {\tt DemoContents} 
in the subdirectory {\tt demos}:
\begin{enumerate}
\item three-point functions
\item four-point functions
\item five-point functions
\item six-point functions
\item calculation of 4-photon helicity amplitudes
\item numerical stability demo: $\det G\to 0$
\item numerical stability demo: $\det S\to 0$
\item Golem $\leftrightarrow$ LoopTools conventions
\end{enumerate}
The items above contain the following options:
\begin{itemize}
\item Three-point functions:
\begin{enumerate}
  \item one off-shell leg
  \item two off-shell legs
  \item three off-shell legs\\
  For each of the three options above, one can choose to calculate:
\begin{enumerate}
  \item scalar three-point function in n dimensions
  \item three-point function in n dimensions with one Feynman parameter $(z_1)$ in the numerator
  \item three-point function in n dimensions with two Feynman parameters $(z_1\,z_2)$
  \item three-point function in n dimensions with three Feynman parameters $(z_1^2\,z_3)$
  \item scalar three-point function in n+2 dimensions
  \item three-point function in n+2 dimensions with one Feynman parameter $(z_2)$
\end{enumerate}    
\end{enumerate}
\item Four-point functions:
\begin{enumerate}
    \item no off-shell leg
    \item one off-shell leg
    \item two opposite off-shell legs
    \item two adjacent off-shell legs
    \item three off-shell legs
    \item four off-shell legs\\
  For each of the five options above, one can choose to calculate:
\begin{enumerate}
  \item scalar four-point function in n dimensions
  \item four-point function in n dimensions with one Feynman parameter $(z_1)$
  \item four-point function in n dimensions with two Feynman parameters $(z_1\,z_4)$
  \item four-point function in n dimensions with three Feynman parameters $ (z_1^2\,z_3)$
  \item four-point function in n dimensions with four Feynman parameters $(z_1\,z_2\,z_3\,z_4)$
  \item scalar four-point function in n+2 dimensions
  \item four-point function in n+2 dimensions with two Feynman parameters $(z_1\,z_2)$
  \item scalar four-point function in n+4 dimensions
\end{enumerate}      
\end{enumerate}    
\item Five-point functions:
\begin{enumerate}
 \item form factor for five-point function, rank 0
 \item form factor for five-point function, rank 3 ($z_1\,z_2\,z_4$ in numerator)
 \item form factor for five-point function, rank 5 ($z_1\,z_2\,z_3\,z_4\,z_5$ in numerator)
 \item form factor for a pinched 5-point diagram (propagator 3 missing), rank 0
 \item form factor for a doubly pinched 5-point diagram (propagators 1 and 4 missing), rank 0
\end{enumerate}    
\item Six-point functions:
\begin{enumerate}
  \item form factor for six-point function, rank 0
  \item form factor for six-point function, rank 4 ($z_1^2\,z_2\,z_3$ in numerator)
  \item form factor A5 for pinched diagram, propagator 3 missing, rank 0
  \item form factor for double pinched diagram, propagators 2,5 missing, rank 0
  \item form factor for triple pinched diagram, propagators 2,4,6 missing, rank 0
\end{enumerate}        
\item Calculation of 4-photon helicity amplitudes: \\
  the purpose of this example is to demonstrate how to use {\tt golem95} for the 
  calculation of full amplitudes. 
  It calculates  three different helicity configurations of the 
  on-shell 4-photon amplitude for a certain kinematic point.
\item Numerical stability demo: $\det G\to 0$:\\
   calculates a rank three four-point function (in 6 dimensions) 
   in a region where $|B|=\det G/\det S$ becomes small, 
   i.e. where a representation based on the reduction to scalar integrals would fail.
   The Feynman parameters in the numerator are $z_1\,z_2^2$.
   The example follows closely the one described in section 7.2 of \cite{Binoth:2005ff}
   and is also described in the golem95 manuscript:
   The program makes 30 iterations where $B=-\det G/\det S$ becomes smaller in 
   each  iteration. 
   The results for real and imaginary parts of $I_4^6(z_1\,z_2^2)$
   are written to the file {\tt demo\_detG.dat} as a function of $x$, 
   where $ |B|~x^2$ for small $x$.
   The files {\tt plotDetG\_Re.gp} and {\tt plotDetG\_Im.gp} can be used to 
   plot the result with gnuplot by {\tt load 'plotDetG\_Re/Im.gp' }.
   One can see from the plots that 
   The file {\tt demo\_detG.txt} contains the details of the kinematics 
   for each iteration.
\item Numerical stability demo: $\det S\to 0$:\\ 
    tests the rank 5 five-point tensor coefficient $A^{5,5}(1,1,1,1,1)$ 
  with respect to its behaviour 
  when a sub-determinant $\det S \sim (\det G)^2 \to 0$.
  The results for real and imaginary parts of the $\epsilon^0$ part of $A^{5,5}$
  are written to the file {\tt demo\_a55\_dets\_sing.dat} as a function 
  of the transverse momentum of particle 5
  and can be plotted with gnuplot by {\tt load 'plot\_demo\_A55.gp'}.
\item  Relation between Golem output and LoopTools format:\\
       produces Golem output for four-point functions up to rank four 
       and gives the relation to LoopTools conventions.
       If LoopTools is linked, the lines containing the call of LoopTools 
       functions can be uncommented to produce LoopTools output in parallel.
\end{itemize}


\end{document}